\documentclass[a4paper,11pt]{article}
\usepackage{jheppub} 
\usepackage{lineno}

\title{\boldmath Holographic reconstruction of flat spacetime}

\author{Zezhuang Hao}
\affiliation{School of Mathematical Sciences, \\Highfield, University of Southampton,\\
SO17 1BJ Southampton, UK}

\emailAdd{Z.Hao@soton.ac.uk}

\abstract{The flat/CFT dictionary between the bulk gravitational theory and boundary conformal field theory is systematically developed in this paper. Asymptotically flat spacetime is built up by asymptotically AdS hyperboloid slices in terms of Fefferman Graham coordinates together with soft modes propagating between different slices near the null boundary. Then we construct the flat holography dictionary based on studying Einstein equation at zero and first order and it turns out that these correspond to the description of hard and soft sector for the field theory from the boundary point of view. The explicit expression for energy-stress tensor is also determined by performing holographic renormalisation on the Einstein Hilbert action. By studying the anomalies of the energy-stress tensor, we obtain the leading and subleading contribution to the central charge.   Einstein equations in the bulk are related to the Ward identities of the boundary theory and we find that the boundary CFT energy-stress tensor is not conserved due to the existence of radiative soft modes which will generate the energy flow through the null boundary. }

\begin{document}
\maketitle
\flushbottom
\section{Introduction}

The AdS/CFT correspondence \cite{Maldacena:1997re,Witten:1998qj,Gubser:1998bc} has been brought up for long time while there are lots of attempts \cite{deHaro:2000wj,deBoer:2003vf,Mann:2005yr,Costa:2012fm,Strominger:2017zoo,Ball:2019atb,raclariu2021lectures,Pasterski:2021rjz,Nguyen:2021ydb} aiming to construct the duality relation between the quantum gravity theory with (asymptotically) flat background and the boundary conformal field theories. It is not clear whether the flat/CFT could work in the same way as AdS/CFT or not until recently the construction of flat/CFT dictionary from the bottom-up point of view \cite{Hao:2023wln}. Basic principles and underlying physical interpretations for flat holography dictionary are discussed using the scalar field as an example. Here, following the same spirit, we will establish the dictionary between the $d+2$ dimensional bulk theories including gravity and the $d$ dimensional boundary conformal field theories on the celestial sphere thus make the construction of flat/CFT duality more complete and concrete.

To study the gravitational theory on asymptotically flat background in a formal way, the first step is to specify the definition of asymptotic flatness. The general guideline for defining asymptotic flatness is that it must make the spacetime close enough to flat case while the deformations should at the same time contain enough non trivial physical contents for us to investigate. For example, the well studied Bondi gauge \cite{Bondi:1962px,Sachs:1962wk,Sachs:1962zza} in gravity literature. To establish the flat/CFT dictionary, one also needs to construct the map of data between the bulk and boundary but there is no clue on how to decompose the bulk data in Bondi gauge. Noting that Minkowski spacetime is foliated by AdS hyperboloids and the specific map of data for AdS/CFT is clearly studied in terms of Fefferman Graham coordinates \cite{AST_1985__S131__95_0,Graham:1999jg,Fefferman:2007rka,Henningson:1998gx,deHaro:2000vlm,Balasubramanian:1999re}, therefore in this paper we choose to define the asymptotically flat spacetime in terms of Fefferman Graham like coordinates written as
\begin{eqnarray}
    ds^2&=&-d\tau^2+\tau^2\left(\frac{d^2\rho}{4\rho^2}+\rho g_{ij}(\rho,x)dx^idx^j\right)\nonumber\\
    &&+\frac{\tau}{\rho^2}m(\rho,x)d^2\rho+\tau\rho\;\sigma_{ij}(\rho,x)dx^idx^j+ \tau A_i(\rho,x)d\rho dx^i+\cdots,\\
 \nonumber
\end{eqnarray}
where the first line is the leading contribution to the spacetime coming from the asymptotic AdS slices and the second line is the subleading contribution for large $\tau$. We will see that most of the non trivial physical results in addition to the AdS/CFT duality would come from the existence of such subleading sector. In section \ref{coordinates}, we will explore such gauge in a careful way by determining the asymptotic symmetries and solving Einstein equation at different order of $\tau$ and $\rho$. The strategy here is that we choose to expand the parameters $g_{ij}(\rho,x),m(\rho,x),\sigma_{ij}(\rho,x),A_i(\rho,x)$ in terms of $1/\rho$ and determine the constraints between the coefficients $g^{(2k)}(x), m^{(2k)}(x), \sigma^{(2k)}(x),A_i^{(2k)}(x)$ by solving the Einstein equations $R_{\mu\nu}=0$ at the zero and first order of $1/\tau$.

After the study of asymptotically flat spacetime, for $d=2$, we propose the flat/CFT dictionary as 
   \begin{equation}
    {\rm exp}\;\big(i S_{{\rm gr,ren}}[G]\big)=\Big\langle {\rm exp } \;\frac{1}{2}\int_{S^2} d^2x \sqrt{\bar{G}}\;\bar{G}^{ij}\;T_{ij}\Big\rangle
\end{equation}
where $G$ and $\bar{G}$ are the bulk metric and the background metric for the CFT, respectively. $T_{ij}$ is the energy-stress tensor of the boundary CFT. To make such dictionary well defined and work the same way as the AdS/CFT dictionary, one further needs to perform proper renormalisation procedure on the bulk gravitational action $S_{\rm gr}[G]$ making it finite and to specify the exact map of data between two sides. Here the exact map means that given the bulk metric we should be able determine boundary data $\bar{G}$ and $T_{ij}$ or vice versa. These are two main obstacles during the development of flat holography dictionary and we will discuss them in section \ref{Dictionary}. 

In the context of AdS/CFT, one needs to perform holographic renormalisation in order to obtain the finite renormalised action $S_{\rm gr,ren}[G]$. The infinity coming from the integral over the whole spacetime is treated as the IR divergence and is regulated by choosing the AdS spatial radius $\rho$ as the IR cut off. Here for flat spacetime, we have one more timelike non-compact direction labeled by $\tau$ and we simply choose to impose a bound $L$ on it, i.e $\tau\in [0,L]$.  After performing the holographic renormalisation in the given interval $[0,L]$, we obtain the renormalised gravitational action and then further propose the map of data between bulk and boundary as
\begin{equation}
    \bar{G}_{ij}= g^{(0)}_{ij}+\frac{1}{L}\sigma_{ij}^{(0)}+\cdots,
\end{equation}
where $\bar{G}_{ij}=g^{(0)}_{ij}$ is originally the AdS/CFT map and now we are taking the soft sector into consideration treating $L$ as the energy cut-off of the boundary CFT. Using the flat/CFT dictionary, we obtain a specific expression for the energy stress tensor $\langle T_{ij}\rangle$, given by
\begin{eqnarray}
   && \langle T_{ij} \rangle=\frac{iL^2}{16\pi G_N}(g_{(2)ij}-g_{(0)ij}{\rm Tr}g_{(2)})- \frac{iL}{8\pi G_N}\Big(({\rm Tr}\;\sigma^{(2)}-\sigma^{(0)}_{kl}\;g_{(2)}^{kl})\;g^{(0)}_{ij}\\ &&\qquad\qquad\qquad\qquad\qquad\qquad+{\rm Tr}g^{(2)}\sigma_{ij}^{(0)}-\sigma^{(2)}_{ij}+2m^{(0)}(g^{(2)}_{ij}-g^{(0)}_{ij}{\rm Tr} g^{(2)})\Big)+\cdots.\nonumber
\end{eqnarray}
Moreover, by considering the anomalies of the energy-stress tensor $\langle T^i_i \rangle$, the central charge is then determined to be 
\begin{equation}
     c=\frac{i3L^2}{4G_N}+\frac{iML3(\alpha-2)}{2G_N}
\end{equation}
where $M=m^{(0)}$ and $\alpha$ are constants characterising the behaviour of asymptotically flat spacetime. The behaviour of the leading term is already argued in the work \cite{Cheung:2016iub,Pasterski:2022lsl,Ogawa:2022fhy} and here we determine its value in a precise way. With the help of flat/CFT dictionary, we can in principle determine all the subleading contributions and in this paper we just present the first order term. 

The Einstein equations will impose constraints on the parameters in the metric and those constrains on the boundary will correspond to the Ward identity of the stress tensor. The story for AdS/CFT or for the hard sector is that the bulk equation of motion implies that the conservation of energy-stress tensor $\nabla^j\langle T_{ij}\rangle=0$. By checking the Einstein equations at the first order, we find that it will not give us the conservation of the soft energy-stress tensor while we have
\begin{equation}
    \bar{\nabla}^j\langle T_{ij} \rangle=\frac{iL}{16\pi G_N}(8\nabla_i m^{(2)}+4M\nabla_i{\rm Tr}g^{(2)}-\sigma^{mk}_{(0)}\nabla_m g^{(2)}_{ki}),
\end{equation}
where ${\bar{\nabla}}_i, \nabla_i$ are the covariant derivative with respect to $\bar{G}_{ij}, g^{(0)}_{ij}$, respectively.  Thus one can interpret the soft modes as the radiation modes which generate flow of energy at the null boundary leading to a non conserved energy-stress tensor from the boundary point of view.
\section{Fefferman Graham Gauge} \label{coordinates}
In this section, we will start from the coordinate for asymptotically flat spacetime then recast it to the asymptotically AdS form. For the spacetime of dimension $d+2$, we first introduce the standard coordinate $X^\mu$ for $\mu =0,1, \cdots, d+1$ in which the flat metric takes the form $\eta_{00}=-1$, $\eta_{11}=\cdots \eta_{d+1,d+1}=1$. The Euclidean AdS of dimension $d+1$ can be regarded as the hyperboloid embedded in the $d+2$ flat spacetime given by the relation \footnote{Such patch is called Milne wedge for Minkowski while one can study Rindler wedge by the analytic continuation of the radius $\tau\rightarrow i\tau$. The Rindler wedge is sliced by dS hypersurfaces.}
\begin{equation}
    -(X^0)^2+(X^1)^2+\cdots+(X^{d+1})^2=-\tau^2,
\end{equation}
in which $\tau\geq 0$ is the radius of the AdS surface. 

Motivated by such foliation, we now choose to write the asymptotically flat spacetime  given by the metric $G(X)$ into the form 
\begin{equation*}
    ds^2=G_{\mu\nu}(X)dX^\mu dX^\nu=-d\tau^2+\tau^2 \hat{G}_{ab}(\tau,y)dy^a dy^b,\label{gauge}
\end{equation*}
where $a,b=1,\cdots , d+1$ and we have fixed the gauge in order to make the $d\tau dy^a$ term vanish. One can always manage to find such gauge by performing local diffeomorphism transformation. For flat spacetime, the metric $\hat{G}_{ab}$ will be independent of $\tau$ and reduced to the metric for $AdS_{d+1}$ described by $y$ while it will become asymptotically AdS when the spacetime is near flat. At the null boundary, we can expand the metric $\hat{G}_{ab}(\tau,y)$ around $1/\tau=0$ written as 
\begin{equation}
    \hat{G}_{ab}(\tau,y)=\hat{G}^{0}_{ab}(y)+\hat{G}^{1}_{ab}(y)\frac{1}{\tau}+\cdots=\hat{G}^{0}_{ab}+h_{ab}(\tau,y), \label{pert}
\end{equation}
in which $\hat{G}^{0}_{ab}(y)$ is the metric for $d+1$ dimensional asymptotically AdS while $\hat{G}^{n}_{ab}(y)$ are higher order corrections and they together form a complete description of asymptotically flat spacetime. We put all the higher order corrections into the term $h_{ab}$. In this article, to study asymptotically AdS spacetime $G^{0}_{ab}$, we choose to use Fefferman and Graham coordinates written as 
\begin{eqnarray}
   \hat{ G}^{0}_{ab}dy^a dy^b&=&\frac{1}{r^2}(dr^2+g_{ij}(r,x)dx^idx^j)\\
    &=& \frac{d\rho^2}{4\rho^2}+\rho\; g_{ij}(x,\rho)dx^idx^j,
\end{eqnarray}
in which $y=(r,x^i)$ for $i=1,\cdots d$ are coordinates on the AdS hyperboloid and $\rho=1/r^2$. The $d$ dimensional metric $g_{ij}(x,\rho)$ has been extensively studied in the AdS/CFT literature and the main method is that one can organise the data by doing expansion of the order $\rho$ for $\rho\rightarrow \infty$. The leading term will contribute to the AdS spacetime while lower order terms are asymptotically AdS corrections.
\subsection{Asymptotic Symmetries}
To illustrate the spacetime structure introduced above in a more precise way, for simplicity, we  take $d=2$ as an example therefore the spacetime becomes asymptotically Minkowski. In this case, the metric is then given by
\begin{eqnarray}
    ds^2&=&-d\tau^2+\tau^2\left(\frac{d^2\rho}{4\rho^2}+\rho\; dzd\bar{z}+\cdots\right)\nonumber\\
    &&+\frac{\tau}{\rho^2}m(\rho,z,\bar{z})d^2\rho+\tau\rho\;\sigma_{z\bar{z}}(\rho,z,\bar{z})dzd\bar{z}+2 \tau A_z(\rho,z,\bar{z})d\rho dz+\tau\rho\; \sigma_{zz}(\rho,z,\bar{z})dzdz \label{asmetric}\nonumber\\
    &&+\;{\rm c.c}\;+\;\cdots, \label{casymetric}
\end{eqnarray}
in which the Minkowski space is written in terms of Milne coordinates $(\tau,\rho,z,\bar{z})$ at the first line while the first order deviations with respect to $\tau$ are described by the functions $m(\rho,z,\bar{z})$, $A_z(\rho,z,\bar{z})$, $\sigma_{z\bar{z}}(\rho,z,\bar{z})$,  $\sigma_{zz}(\rho,z,\bar{z})$ together with the corresponding complex conjugates. The dots in the first line represent the asymptotically AdS deformation while the dots in the third line represent higher order contributions according to $1/\tau$. Written in the form of $G_{\mu\nu}$, the metric are
\begin{equation}
    G_{\rho\rho}=\frac{\tau^2}{4\rho^2}+\frac{\tau}{\rho^2}m+\mathcal{O}(\tau^0)
\end{equation}
for $\rho\rho$ component and for $z\bar{z}$ component we have
\begin{equation}
    G_{z\bar{z}}=\frac{\tau^2\rho}{2}+\tau\rho\; \sigma_{z\bar{z}}+\mathcal{O}(\tau^0).
\end{equation}
At leading order, $G_{\rho z}$ $G_{zz}$ will vanish while for asymptotic flat spacetime there could be subleading contribution like
\begin{equation}
    G_{\rho z}=\tau A_z+\mathcal{O}(\tau^0)\qquad  G_{\rho \bar{z}}=\tau A_{\bar{z}}+\mathcal{O}(\tau^0)
\end{equation}
and 
\begin{equation}
    G_{zz}=\tau\rho\; \sigma_{zz}+\mathcal{O}(\tau^0)\qquad  G_{\bar{z}\bar{z}}=\tau \rho \;\sigma_{\bar{z}\bar{z}}+\mathcal{O}(\tau^0).
\end{equation}
Now, we will study asymptotic symmetry which is the diffeomorphism that preserves the metric of the form  shown in \eqref{casymetric}. Given the Killing vector $\xi(\tau,\rho,z,\bar{z})$, the variation of the metric $\delta G_{\tau\rho}$ is then deduced to be
\begin{eqnarray}
    \delta G_{\tau\rho}&=&-\partial_\rho \xi^\tau +\frac{\tau^2}{4\rho^2}\partial_\tau \xi^\rho \nonumber\\
    && + \frac{\tau}{\rho^2} m \partial_\tau \xi^\rho+\tau A_z\partial_\tau \xi^z+\tau A_{\bar{z}}\partial_\tau \xi^{\bar{z}}+\mathcal{O}(\tau^0)\label{vtr}
\end{eqnarray}
and for $\delta G_{\tau z}$ we have
\begin{eqnarray}
    \delta G_{\tau z}&=& -\partial_z \xi^\tau+\frac{\tau^2 \rho}{2}\partial_\tau \xi^{\bar{z}}\nonumber\\
    && +\tau A_{z}\partial_\tau \xi^\rho+\tau\rho\;\sigma_{zz}\partial_\tau \xi^z+\tau\rho \;\sigma_{z\bar{z}}\partial_\tau \xi^{\bar{z}} +\mathcal{O}(\tau^0),\label{vtz}
\end{eqnarray}
in which we have shown the variation up to $\mathcal{O}(\frac{1}{\tau})$ and higher order terms are omitted. For other terms we have
\begin{equation}
    \delta G_{z\bar{z}}=\frac{\tau^2}{2}\xi^\rho+\frac{\tau^2\rho}{2}(\partial_z\xi^z+\partial_{\bar{z}}\xi^{\bar{z}})+\mathcal{O}(\tau), 
\end{equation}
\begin{equation}
    \delta G_{zz}=\tau^2 \rho\;\;\partial_z \xi^{\bar{z}}+\mathcal{O}(\tau),
\end{equation}
\begin{equation}
    \delta G_{\rho z}=\frac{\tau^2}{4\rho^2}\partial_z \xi^\rho+\frac{\tau^2\rho}{2}\partial_\rho \xi^{\bar{z}}+\mathcal{O}(\tau),
\end{equation}
\begin{equation}
    \delta G_{\rho\rho}=\frac{\tau^2}{2\rho^3}(\rho\partial_\rho \xi^\rho-\xi^\rho)+\mathcal{O}(\tau).\label{vrr}
\end{equation}
To find out the asymptotic symmetries, one need to determine the Killing vector $\xi$ which can be expanded as
\begin{equation}
    \xi^\mu(\tau,\rho,z,\bar{z})=\xi^\mu_{0}(\rho,z,\bar{z})+\frac{1}{\tau}\xi^\mu_{1}(\rho,z,\bar{z})+\cdots
\end{equation}
where $\xi ^\mu_k$ for $k\in N$ are coefficients associated to the term $1/\tau^k$. For $\delta G_{\tau\rho}$ and $\delta G_{\tau z}$ they have to vanish since we are working in the gauge $G_{\tau\rho}=G_{\tau z}=0$. From the expression \eqref{vtr} and \eqref{vtz}, at the order of $\mathcal{O}(1)$ we obtain
\begin{equation}
    \partial_z\xi^\tau_{0}+\frac{\rho}{2}\xi^{\bar{z}}_{1}=0\qquad\partial_\rho \xi^\tau_{0}+\frac{1}{4\rho^2}\xi^\rho_{1}=0.
\end{equation}
Moreover, we have 
\begin{equation}
    A_z\xi^\rho_{1}+\rho\; \sigma_{zz}\xi^z_{1}+\rho \;\sigma_{\bar{z}\bar{z}}\xi^{\bar{z}}_{1}=0
\end{equation}
and
\begin{equation}
    m \;\xi^\rho_{1}+\rho^2 A_z \xi^z_{1}+\rho^2A_{\bar{z}}\xi^{\bar{z}}_{1}=0
\end{equation}
when considering the $\delta G_{\tau\rho}=\delta G_{\tau z}=0$ of the order $\tau^2$. For $\delta G_{zz}$ and $\delta G_{\bar{z}\bar{z}}$, the contribution at the order $1/\tau$ should vanish in order to preserve the condition of asymptotic flatness therefore we have
\begin{equation}
      \partial_z \xi^{\bar{z}}_{0}=\partial_{\bar{z}}\xi^z_{0}=0.
\end{equation}
For the same reason we have
\begin{equation}
 \rho \partial_\rho \xi^\rho_{0}-\xi^\rho_{0}=0
\end{equation}
when considering the component $G_{\rho\rho}$ in the \eqref{vrr}.
Then the killing vector can be written as 
\begin{eqnarray}
   && \xi^\tau=\chi(z,\bar{z}), \\
   && \xi^\rho=  0,           \\
   && \xi^z=Y^z(z,\bar{z}), \\
   && \xi^{\bar{z}}=Y^{\bar{z}}(z,\bar{z}),
   \end{eqnarray}
where $\chi$ is an arbitrary function on $z,\bar{z}$ and we have $\partial_z Y^{\bar{z}}=\partial_{\bar{z}}Y^z=0$ therefore the BMS group \cite{Bondi:1962px,Sachs:1962wk,Sachs:1962zza} is recovered at leading order. The transformation of the spacetime metric under such symmetry group is then given by
\begin{eqnarray}
    \delta g_{z\bar{z}}&=& g_{z\bar{z}}(\partial_zY^z+\partial_{\bar{z}}Y^{\bar{z}}) +Y^z\partial_z g_{z\bar{z}} +Y^{\bar{z}}\partial_{\bar{z}}g_{z\bar{z}}\\
    \delta g_{zz}&=& 2g_{zz}\;\partial_z Y^z+Y^z\partial_z g_{zz}+Y^{\bar{z}}\partial_{\bar{z}}g_{zz}\\
    \delta A_z&=& Y^z\partial_z A_z+Y^{\bar{z}}\partial_{\bar{z}}A_z +A_z \partial_z Y^z\\
    \delta \sigma_{zz} &=& Y^z\partial_z \sigma_{zz}+ Y^{\bar{z}}\partial_{\bar{z}}\sigma_{zz}+2\sigma_{zz}\partial_z Y^z+2 \chi g_{zz}\\
    \delta \sigma_{z\bar{z}}&=&Y^z\partial_z \sigma_{z\bar{z}}+ Y^{\bar{z}}\partial_{\bar{z}}\sigma_{z\bar{z}}+(\partial_zY^z+\partial_{\bar{z}}Y^{\bar{z}}) \sigma_{z\bar{z}}+2\chi g_{z\bar{z}}
\end{eqnarray}
where we have used the complex metric $g_{z\bar{z}}$ for the hard sector. From above translation rules, one can see that the superrotation part described by $Y^z$, $Y^{\bar{z}}$ will act on the leading and subleading part of the metric while the supertranslation part described by $\chi$ will only act on the subleading part.
\subsection{Equation of Motion}
Now we turn to study the dynamics of the gravitational system. The general $d+2$ dimensional asymptotic flat spacetime is described by the coordinate
\begin{eqnarray}
    ds^2&=&-d\tau^2+\tau^2\left(\frac{d^2\rho}{4\rho^2}+\rho g_{ij}(\rho,x)dx^idx^j\right)\nonumber\\
    &&+\frac{\tau}{\rho^2}m(\rho,x)d^2\rho+\tau\rho\;\sigma_{ij}(\rho,x)dx^idx^j+ \tau A_i(\rho,x)d\rho dx^i+\cdots \label{asmetric}\\
 \nonumber
\end{eqnarray}
in terms of the real coordinates $(\tau,\rho,x^i)$ for $i=1,\dots, d$. In the first line, the metric is mainly built up from asymptotically AdS hyperboloids. It describes the hard sector of the gravitational theory and manifests the superrotation symmetry. The second line is the subleading contribution to the asymptotically flat spacetime according to $1/\tau$. It describes the soft modes coming from the radiation and manifests the superrotation and supertranslation symmetry.

In this section, we are going to determine the constraints on the metric $g_{ij}$ and the soft parameters by checking the Einstein equations $R_{\mu\nu}=0$ at different orders. Starting with the connections, they are given by 
\begin{equation}
    \Gamma^{\tau}_{ab}=\tau \hat{G}_{ab}+\frac{1}{2}\tau^2\partial_\tau \hat{G}_{ab}
\end{equation}
and
\begin{eqnarray}
    \Gamma^{a}_{b\tau}=\frac{\delta^a_b}{\tau}+\frac{1}{2}\hat{G}^{ac}\partial_\tau \hat{G}_{cb}
\end{eqnarray}
while we have $\Gamma^{\tau}_{\tau\tau}=\Gamma^{\tau}_{\tau a}=\Gamma^a_{\tau\tau}$=0 and $\Gamma_{bc}^a=\hat{\Gamma}_{bc}^a[\hat{G}]$.
In our definition the Ricci tensor is now given by 
\begin{equation}
    R_{ab}[G]=R^{d+1}_{ab}[\hat{G}]-\Gamma_{ab}^\tau \Gamma^c_{c\tau}+2\Gamma^c_{a\tau}\Gamma^\tau_{bc}-\partial_\tau\Gamma^\tau_{ab},
\end{equation}
which can be further decomposed into
\begin{eqnarray}
    R_{ab}[G]&=&R^{d+1}_{ab}[\hat{G}]-d\hat{G}_{ab}\\
    &&-\frac{d+1}{2}\tau \partial_\tau\hat{ G}_{ab}-\frac{1}{2}\tau^2\partial^2_\tau \hat{G}_{ab}-\frac{\tau}{2}\hat{G}_{ab}\hat{G}^{cd}\partial_\tau \hat{G}_{cd}\\
    &&+\frac{1}{2} \tau^2 \partial_\tau \hat{G}_{cb}\partial_\tau \hat{G}_{da} \hat{G}^{cd}. \label{ab}
\end{eqnarray}
For $\tau\tau$  component we have 
\begin{equation}
    R_{\tau\tau}[G]=\frac{1}{\tau}\hat{G}^{ab}\partial_\tau \hat{G}_{ab}+\frac{1}{2}\partial_\tau(\hat{G}^{ab}\partial_\tau \hat{G}_{ab})+\frac{1}{4}\hat{G}^{ac}\hat{G}^{bd}\partial\tau \hat{G}_{ab}\partial_\tau \hat{G}_{cd} \label{tt}
\end{equation}
while for $\tau a$ one can deduce that
\begin{equation}
    R_{ a \tau}[G]=\frac{1}{2}\hat{\nabla}_a(\hat{G}^{bc}\partial_\tau \hat{G}_{bc})-\frac{1}{2}\hat{\nabla}_b(\hat{G}^{bc}\partial_\tau \hat{G}_{ca}), \label{ta}
\end{equation}
where the covariant derivative is with respect to the metric $\hat{G}_{ab}$. In practice, to study the equation of motion at different orders, we choose to write the Ricci curvature perturbatively according to the expansion \eqref{pert}, which means we have
\begin{equation}
    R_{\mu\nu}[G^{0}+h]=R^{0}_{\mu\nu}[G^{0}]+\frac{1}{\tau}R^{1}_{\mu\nu}+\cdots
\end{equation}
where the zero order mainly comes from the hard sector of the metric described by $G^{0}$ and the soft modes will go into the first or higher order terms.
\subsubsection{Zero Order}
For convenience, we will denote $\hat{G}^{0}_{ab}(y)$ as $\hat{G}_{ab}(y)$ in this subsection and consider the equation of motion at leading order of $1/\tau$. In such case, connections involving $\tau$ component are given by $\Gamma^\tau_{ab}=\tau \hat{G}_{ab}, \Gamma^a_{b\tau}=\frac{1}{\tau}\delta^a_b \label{tab}$ and $\Gamma^{\tau}_{\tau\tau}=\Gamma^\tau_{\tau a}=\Gamma^a_{\tau\tau}=0$.
For the connections not involving $\tau$ component denoted as $\Gamma^a_{bc}$, they are given by the direct $d+1$ dimensional calculation using the AdS metric $\tau^2 \hat{G}_{ab}(y)$. Now, to solve the vacuum Einstein equation with zero cosmology constant
\begin{equation}
    R_{\mu\nu}[G]=0,\label{Ein}
\end{equation}
we should deduce the Ricci curvature $R$ written as $R[G]=R^a_a[G]+R^\tau_\tau[G]$. One can easily verify that $R_{\tau\mu}=0$, while for $R_{ab}$ one has 
\begin{equation}
    R_{ab}[G]= R^{(d+1)}_{ab}[\hat{G}]-d\hat{G}_{ab},
\end{equation}
in which we have introduced the notion $R^{(d+1)}[\hat{G}]$ to denote that the Ricci curvature induced on the $d+1$ dimensional AdS hyperboloid. Therefore, the Ricci curvature for near flat spacetime is then deduced to be
\begin{equation}
    R[G]= R^{(d+1)}[\hat{G}]-{d(d+1)}
\end{equation}
and the Einstein equation in \eqref{Ein} is equivalent to 
\begin{equation}
    R^{(d+1)}_{ab}[\hat{G}]-\frac{1}{2}R^{(d+1)}[\hat{G}]\;\hat{G}_{ab}=\Lambda \hat{G}_{ab}\label{adEin}
\end{equation}
for
\begin{equation}
    \Lambda=-\frac{d(d-1)}{2}.
\end{equation}
We can treat $\Lambda$ as the effective cosmology constant and $\tau$ as the effective AdS radius since we are recasting the curvature $R[G]$ for $d+2$ dimensional near flat metric $\hat{G}_{\mu\nu}$ into the induced  curvature $R^{(d+1)}[\hat{G}]$ for $d+1$ dimensional asymptotically AdS metric $\hat{G}_{ab}(y)$.

In terms of Fefferman Graham gauge, the study of equation of motion at zero order is equivalent to the study of the differential equation for $g_{ij}(x,\rho)$. The function $g_{ij}(x,\rho)$ are determined by \cite{Graham:1999jg,Fefferman:2007rka,Henningson:1998gx,deHaro:2000vlm}
\begin{eqnarray}
    &&\rho^2(2\rho g''_{ij}+4g'_{ij}-2\rho g^{lm}g'_{mj}g'_{li}+\rho g^{lm}g'_{lm}\;g'_{ij})+R_{ij}[g]+(d-2)\rho^2g'_{ij}+\rho^2g^{lm}g'_{lm}\;g_{ij}=0\nonumber\\
    &&\nabla_i (\; g^{lm}g'_{lm})-\nabla^jg'_{ij}=0 \label{equation}\\
    &&g^{ij}(\rho g''_{ij}+2g'_{ij})-\frac{1}{2}\rho g^{ik}g^{jm}g'_{ij}g'_{km}=0\nonumber
\end{eqnarray}
according to the equation \eqref{Ein} or \eqref{adEin} and the covariant derivative is with respect to the metric $g_{ij}$. A brief study of such equation is shown in appendix \ref{ads} and here we have $g'_{ij}=\partial_\rho g_{ij}$. Moreover, for even $d$, one has the expansion 
\begin{equation}
    g_{ij}(x,\rho)=g^{(0)}_{ij}+\rho^{-1}g^{(2)}_{ij}+\cdots +\rho^{-d/2}g^{(d)}_{ij}+c_{ij}\rho^{-d/2}{\rm log}\rho+\cdots \label{expan}
\end{equation}
when $\rho$ goes to infinity. Here, following the convention from previous literature, the superscript $2k$ in the coefficients $g^{(2k)}_{ij}$ are used to keep track of the order of $r$. Or equivalently, $k$ is used to keep track of the order of $\rho$. Coefficients $g^{(2k)}_{ij}$ are uniquely determined by the lower order terms via checking the equation \eqref{equation} at the order of $\rho^k$ while such procedure will fail until $2k=d$. In such case, the equation of motion will only allow us to determine ${\rm Tr }g^{(d)}$ and it also leaves us freedom to introduce the traceless algorithm term parameterised by $c_{ij}$. Fox example, by checking the first two equations at leading order, one can obtain the relation
\begin{equation}
    R_{ij}[g^{(0)}]=(d-2)g^{(2)}_{ij}+g^{(0)}_{ij}{\rm Tr}g^{(2)}\label{relation1}
\end{equation}
together with
\begin{equation}
    \nabla_i({\rm Tr}g^{(2)})-\nabla^j g^{(2)}_{ji}=0 \label{relation2}
\end{equation}
where the covariant derivative $\nabla_i$ 
 and the trace ${\rm Tr}$ now are with respective to the metric $g^{(0)}_{ij}$ while we will keep such convention in the following part of this paper. One can see that $g^{(2)}_{ij}$ is fully determined as the function of $g^{(0)}_{ij}$ for $d\neq 2$ while only the trace part is fixed when $d=2$.
\subsubsection{First Order}
We have studied $R^{0}_{\mu\nu}$ in the previous section while here we are going to deal with $R^{1}_{\mu\nu}$ therefore determine equation of motion at first order. It is easier to calculate the $R_{\tau\tau}$ and $R_{\tau a}$ components by checking the formula \eqref{tt} and \eqref{ta}. The $R_{\tau\tau}=0$ will be trivial at first order while for $R^1_{\tau a}$, we have
\begin{equation}
    R^1_{\tau a}=\frac{1}{2}\hat{\nabla}_a (\hat{G}_0^{bc}\hat{G}^1_{bc})-\frac{1}{2}\hat{\nabla}^b\hat{G}_{ab}^1
\end{equation}
where the covariant derivative $\hat{\nabla}_a$ is with respect to the metric $\hat{G}^0_{ab}. $ For the $R_{ab}$ components, we have
\begin{equation}
    R^1_{ab}=R^{d+1,1}_{ab}-\frac{d+1}{2}\hat{G}_{ab}^1+\frac{1}{2}\hat{G}_{ab}^1 G^{cd}_0G_{cd}^1 \label{1order}
\end{equation}
from which one can see that the first order contribution $R^{d+1,1}_{ab}$ of  $R^{d+1}_{ab}$ will also contribute to the first order $R^{1}_{ab}$ therefore it will make the results more complicated. The strategy here is to determine the first order term in $R^{d+1}_{ab}$ then add the other terms in \eqref{1order} which also contribute at the first order. 

Such soft sector is described by the first order term $h_{\mu\nu}$. More precisely, it is determined by the parameter $m$, $\sigma_{ij}$ and $A_i$ once the gauge is fixed. To simplify the calculation further, now we consider the expansion of parameter $m$, $\sigma_{ij}$ and $A_i$ by the order $1/\rho$. Taking the parameter $m(\rho,x)$ for example,  we have
 \begin{equation}
     m(\rho,x)=m^{(0)}(x)+\frac{1}{\rho}m^{(2)}(x)+\cdots,
 \end{equation}
 in which $m^{(0)}(x)$ is the leading term while $m^{(2)}(x)$ is the subleading contribution. $m^{(0)}$, $m^{(2)}$ describe the zero order and first order contribution to the soft sector according to the spatial radius of AdS hyperboloid $\rho$. For the parameter $\sigma_{ij}$, $A_i$ we adopt similar convention and the corresponding coefficients are denoted as $A_i^{(2k)}$ and $\sigma^{(2k)}_{ij}$.
 
 Therefore following the equation of motion given by the Einstein equation explicitly showed in the appendix \ref{eom}, we obtain the constraints  
 \begin{equation}
     {\rm Tr}\;\sigma^{(0)}=4d m^{(0)} \label{constraint1}
 \end{equation}
by considering $R_{\rho\rho}=R_{\rho\tau}=0$. Moreover we have
 \begin{equation}
     A^{(0)}_i=0
 \end{equation}
by checking the equation of motion $R_{i\tau}=0$ at the zero order of $1/\rho$. For the equation $R_{ij}=0$, again at leading order, we obtain
 \begin{equation}
     \frac{d-1}{2}\sigma^{(0)}_{ij}+\frac{1}{2}g_{ij}^{(0)}\;{\rm Tr}\;\sigma^{(0)}-2(2d-1)m^{(0)} g_{ij}^{(0)}=0,
 \end{equation}
 which is compatible with the constraint \eqref{constraint1} after taking the trace on both sides. One can also obtain the relation
\begin{equation}
    \nabla^j\sigma^{(0)}_{ji}=4\nabla_im^{(0)} \label{constraint2}
\end{equation}
after acting the covariant derivative $\nabla^j$ on both sides. 

Now we consider the Einstein equations at first order of $1/\rho$ in order to determine $m^{(2)}$, $A^{(2)}_i$ and $\sigma^{(2)}_{ij}$. By checking the equation of motion $R_{i\tau}=0$ at the first order, we have
\begin{equation}
    2\nabla_i m^{(0)}=A^{(2)}_i \label{constraint3}
\end{equation}
while for $R_{\rho i}=0$ we obtain the relation
\begin{equation}
    \frac{3d-1}{2}A^{(2)}_i-d\nabla_im^{(0)}+\frac{1}{4}\nabla^j\sigma^{(0)}_{ij}-\frac{1}{4}\nabla_i{\rm Tr} \sigma^{(0)}=0 .
\end{equation}
Given the above equation, we can deduce that $A^{(2)}_i=0$ after using the relation \eqref{constraint1},\eqref{constraint2} and \eqref{constraint3}. This tells us that the parameter $m^{(0)}$ should be a constant and the zero order coefficient $\sigma_{ij}^{(0)}$ is conserved with respect to the metric $g^{(0)}_{ij}$, written as  
\begin{equation}
    {\rm Tr}\sigma^{(0)}=4dM \qquad \nabla^j\sigma^{(0)}_{ij}=0
\end{equation}
where we have denoted the parameter $m^{(0)}$ as constant $M$. For the equation of motion $R_{\rho\tau}=0$ and $R_{\rho\rho}=0$, they will give us the relations 
\begin{equation}
 dm^{(2)}+\frac{1}{4}{\rm Tr}\;\sigma^{(2)}-{\rm Tr} g_{(2)}m^{(0)}=0 \label{rel0}
\end{equation}
and
\begin{equation}
   \frac{d}{2}m^{(2)}+\frac{1}{8}{\rm Tr}\;{\sigma^{(2)}}-\frac{1}{8}g^{ij}_{(2)}\sigma^{(0)}_{ij}=0.\label{rel1}
\end{equation}
From these two equations one can see that, in order to make $m^{(2)}$ and $\sigma_{ij}^{(2)}$ solvable, one should further impose constraint on $g^{ij}_{(2)}\sigma^{(0)}_{ij}$ thus we have  
\begin{eqnarray}
    g^{ij}_{(2)}\sigma^{(0)}_{ij}=4 {\rm Tr}g^{(2)}m^{(0)}.\label{rel2}
\end{eqnarray}
To obtain $\sigma^{(2)}_{ij}$, one needs to check $R_{ij}=0$ explicitly. The equation is more involved and here we just present the result
\begin{eqnarray}
    && \frac{3-d}{2}\sigma^{(2)}_{ij}+(6-4d)m^{(2)}g^{(0)}_{ij}-6g^{(2)}_{ij}m^{(0)}-\frac{1}{2}({\rm Tr}\sigma^{(2)}-g^{lk}_{(2)}\sigma^0_{lk})g^{(0)}_{ij}\\ &&\qquad\qquad\qquad\qquad+4{\rm Tr}g^{(2)}m^{(0)}g^{(0)}_{ij}-{\rm Tr}g^{(2)}\sigma^{(0)}_{ij}+\frac{1}{2}{\rm Tr}\sigma^{(0)}g^{(2)}_{ij}+\delta R_{ij}=0\nonumber,
\end{eqnarray}
where $R_{ij}[g+\frac{1}{\tau}\sigma]=R_{ij}[g]+\delta R_{ij}$ \footnote{More precisely, we have $\delta R_{ij}=\frac{1}{2}(\nabla^m\nabla_m\sigma_{ij}+\nabla_i\nabla_j\sigma^m_m-\nabla_k\nabla_i\sigma^k_j-\nabla_k\nabla_j\sigma^k_i).$} and we can see that, given the value of $m^{(2)}$, $\sigma^{(2)}_{ij}$ is determined by solving the equation. After taking the trace on both sides, we have
\begin{eqnarray}
    \frac{3-2d}{2}{\rm Tr} \sigma^{(2)}+(6-4d)dm^{(2)} -6{\rm Tr}g^{(2)}\; m^{(0)}+\frac{d}{2}g^{lm}_{(2)}\sigma^{(0)}_{lm}+2dm^{0}{\rm Tr}g^{(2)}=0, \label{rel3}
\end{eqnarray}
from which we will obtain the same relation as \eqref{rel2} after the substitution of \eqref{rel0} or \eqref{rel1}.

From above calculation, we see that the value of $m^{(2k)}$ are related to the trace of soft metric coefficients ${\rm Tr} \sigma^{(2k)}$. More precisely, by considering the constraint \eqref{constraint1}, \eqref{rel0} and \eqref{rel2}, one can find that they obey the more compact relation
\begin{eqnarray}
    {\rm Tr}^{(2k)}(\sigma- 4mg)=0  \label{trace}
\end{eqnarray}
where ${\rm Tr}^{(2k)}$ is defined as the trace over the metric $g^{(2k)}_{ij}$ and we denote ${\rm Tr}^{(0)}={\rm Tr}$. The relation between \eqref{trace} and the Einstein equations is not clear but we expect this is true when going to the higher order. 

Further more, by checking the equation of motion $R_{i\tau}=0$ at the order of $1/\rho^2$, we obtain the relations involving $\nabla^j\sigma_{ij}^{(2)}$ written as
\begin{eqnarray}
    &&(2-d)A_i^{(4)}+2\nabla_im^{(2)}+\frac{1}{2}\nabla_i({\rm Tr}\sigma^{(2)}-g^{kl}_{(2)}\sigma^{(0)}_{kl})\nonumber\\&&\qquad\qquad\qquad\qquad\qquad-\frac{1}{2}\nabla^j\sigma^{(2)}_{ij}+\frac{1}{4}\nabla_kg^{mk}_{(2)}\sigma ^{(0)}_{mi}+\frac{1}{4}\nabla_i g^{mn}_{(2)}\sigma^{(0)}_{mn}=0 \label{constraint4}
\end{eqnarray}
where we have used the relation \eqref{relation2} and the last two terms come from the variation of the connection $\delta \Gamma^i_{jk}$. It turns out that, although a little bit tedious, equation \eqref{constraint4} will be useful for us to study Ward identities of the boundary conformal field theory with the help of flat/CFT dictionary. Moreover, by studying the equation of motion $R_{i\rho}=0$ at second order of $1/\rho^2$ one should be able to determine $A^{(4)}_i$ once $m^{(2)}$ or equivalently ${\rm Tr}\;\sigma^{(2)}$ is fixed.
\section{The flat/CFT Dictionary}\label{Dictionary}

The Einstein-Hilbert action for the gravitational theory on a four dimensional asymptotically flat manifold $M$ with boundary $\partial M$ is given by \cite{Gibbons:1976ue}
\begin{equation}
    S_{{\rm gr}}[G]=\frac{1}{16\pi G_N}\left(\int_M d^{4}X\sqrt{-G}\;R[G]+\int_{\partial M}d^{3}X\sqrt{-\gamma}\;2K\right),
\end{equation}
in which $K$ is the trace of the second fundamental form and $\gamma$ is the induced metric on the boundary. To evaluate the action, we first choose to use the equation of motion \eqref{Ein} and set the boundary as the surface with constant AdS spatial radius $\rho=1/\epsilon$. Then the regulated action is given by
\begin{equation}
    S_{{\rm gr,reg}}=\frac{1}{16\pi G_N}\int_0^\infty d\tau \int_{S^2} d^2 x\sqrt{-\gamma}\;2K\Big|_{\rho=1/\epsilon}
\end{equation}
where we have $\gamma_{\tau\tau}=-1, \gamma_{ij}=\tau^2 \hat{G}_{ij}$.\footnote{In fact, there are three components that belong to the boundary $\partial M$. One is at $\rho=1/\epsilon$ while the other two are at $\tau=0$ and $\tau=\infty$. In this paper we focus on the renormalisation of the divergence at $\rho=1/\epsilon$. For the  integral along the surface of constant $\tau$, we treat them as the assignment of initial and final data. The treatment of the integral at the constant time surface is equivalent to the procedure that we fix the initial modes by hand or inserting proper $i\varepsilon$ description in the path integral. More rigorously, like the treatment for real time holography \cite{Skenderis:2008dg,Skenderis:2008dh,Hao:2023ijx}, one can choose to glue an Euclidean cap at $\tau=0$ surface and make divergence cancelled.}

Moreover, to calculate the integral over $K$, one should notice the relation $K_{ij}=\hat{\nabla}_i n_j$ thus we obtain \footnote{We are abusing the notion here and $\hat{\nabla}$ means the covariant derivative with respect to the metric $\hat{G}_{ij}$. }
\begin{equation}
   \int_0^\infty d\tau\int d^2x \sqrt{-\gamma}\; 2K\Big |_{\rho=1/\epsilon}= \int_0^\infty \tau d\tau \int_{S^2} d^2 x \sqrt{\hat{G}}\frac{\hat{G}^{\rho\mu}}{\sqrt{\hat{G}^{\rho\rho}}}(2\hat{G}^{ij}\partial_i\hat{G}_{j\mu}-\hat{G}^{ij}\partial_\mu \hat{G}_{ij})\Big |_{\rho=1/\epsilon} \label{onshell}
\end{equation}
in which $n^\mu$ is the outward unit normal for the boundary $\partial M$. In our case, for the boundary $\rho=1/\epsilon$, the only non-zero component is $n^{\rho}=\sqrt{G^{\rho\rho}}$. To see the divergent part of the regulated action in a more precise way, for even $d$, one can use the expansion for $g$ shown in  \eqref{expan} and extract the infinite part written as
\begin{equation}
    S_{{\rm gr,reg}}=\frac{1}{16\pi G_N}\int_0^\infty d\tau \tau \int_{S^2}d^2x\;\sqrt{g_{(0)}}\left(\epsilon^{-1}a_{1}(\tau,x)+a_{0}(\tau,x)+{\rm log}\epsilon\; b(\tau,x)+\mathcal{O}(\epsilon^0)\right),\label{reg}
\end{equation}
where $a_i$ and $b$ are the corresponding coefficients. To get the renormalised action $S_{{\rm gr,ren}}$, one should introduce the local and covariant counterterm $S_{{\rm gr,ct}}$ to eliminate the divergence, which takes the form 
\begin{equation}
    S_{{\rm gr,ct}}=\int_{0}^{\infty} d\tau \int _{S^2}d^2x \; f(\tau,z)\sqrt{-\gamma}+\int_{0}^{\infty} d\tau \int _{S^2}d^2x \; g(\tau,z)\sqrt{-\gamma}\; R\;[\gamma]+\cdots
\end{equation}
where $f,g$ are scalar functions of $\tau$, $x$ and they are determined by the coefficients $a_i$, $b$ in \eqref{reg}. Given the renormalised action $S_{{\rm gr,ren}}=S_{{\rm gr,reg}}+S_{{\rm gr,ct}}$ together with the dictionary 
    \begin{equation}
    {\rm exp}\;\big(i S_{{\rm gr,ren}}[G]\big)=\Big\langle {\rm exp } \;\frac{1}{2}\int_{S^2} d^2x \sqrt{\bar{G}}\;\bar{G}^{ij}\;T_{ij}\Big\rangle \label{dictionary}
\end{equation}
the CFT stress tensor $T_{ij}$ is then deduced to be
\begin{equation}
    \langle T_{ij} \rangle=\lim_{\epsilon\rightarrow 0}\frac{2i}{\sqrt{\bar{G}}}\frac{\delta S_{{\rm gr,ren}}}{\;\delta \bar{G}^{ij}},
\end{equation}
where $\bar{G}_{ij}$ is the background metric of the boundary CFT. Before going into the detail of holographic renormalisation, here we briefly discuss the structure of renormalised gravity effective action. Through the renormalisation procedure, the IR divergence is regulated by choosing the AdS spatial radius $\rho=1/\epsilon$ as the low energy cutoff. However, for the flat spacetime, we also have timelike direction labelled by $\tau$ and here we treat it as the UV cutoff from boundary point of view by specifying the range of integral $0\leq\tau\leq L$. Therefore, organized by the powers of $L$, the action takes the form
\begin{equation}
    S_{\rm gr,reg}=S^{0}_{\rm gr,reg}+S^{1}_{\rm gr, reg}+\cdots.
\end{equation}
where $S^0_{\rm gr,ren}$ is the leading contribution while $S^0_{\rm gt,ren}$ is subleading. After performing the integral of $\tau$ during the holographic renormalisation, we have
\begin{equation}
    S_{\rm gr, ren}= L^2 S^{0}_{\rm gr,ren}+ L S^{1}_{\rm gr,ren}+ \mathcal{O}(\log{L}),
\end{equation}
in which $S^{0}_{\rm gr,ren}$ is the  contribution to the renormalised action of the order $L^2$ and $S^{1}_{{\rm gr,ren}}$ is the lower order term. We identify $S^{0}_{\rm gr,ren}$ as the hard sector since it comes from the AdS hyperboloid while the soft sector is identified as $S^{1}_{\rm gr,ren}$ coming from soft modes in the metric.

Given the dictionary \eqref{dictionary}, to perform the calculation and make it work the same as the AdS/CFT dictionary, we need the specific map between the boundary and bulk data and here we propose the relation to be
\begin{equation}
    \bar{G}_{ij}= g^{(0)}_{ij}+\frac{1}{L}\sigma_{ij}^{(0)}+\cdots
\end{equation}
where the boundary background metric is expanded by the order of energy cut off $L$ given by the bulk data $g^{(0)}_{ij}$ and $\sigma^{(0)}_{ij}$.
\subsection{Hard Sector}
In this section, we choose to perform the holographic renormalisation for the hard sector ignoring the soft contribution from $h_{ij}$. It turns out the treatment of the hard sector is equivalent to the linear summation over all the AdS hyperboloid contribution and one can regard this part as the review of AdS/CFT holographic renormalisation. At zero order, the onshell action takes the form
\begin{equation}
      S_{{\rm gr,reg}}=\frac{-1}{16\pi G_N}\int_0^L d\tau \tau \int_{S^2}d^2x\;\sqrt{g}\;2\rho^2\left(\frac{2}{\rho}+g^{ij}\partial_\rho g_{ij}\right)\Big|_{\rho=1/\epsilon}
\end{equation}
and the counterterm is given by
\begin{equation}
    S_{{\rm gr},{\rm ct}}=-\frac{d-1}{8\pi G_N}\int_0^L d\tau \int_{S^2} d^2x \;\frac{1}{\tau}\sqrt{-\gamma}.
\end{equation}
Now, following the above discussion, we will show the result for $d=2$.
In such case, $g_{(0)ij}$ is the metric on the sphere and  the regulated stress tensor $T^{{\rm reg}}_{ij}$ on the celestial sphere is therefore given by  
\begin{equation}
    T^{{\rm reg}}_{ij}=\frac{iL^2}{16\pi G_N}(K_{ij}-\gamma_{ij}K)=\frac{iL^2}{16\pi G_N}\rho\left((d-1)g_{ij}+\rho g_{ij}g^{lk}\partial_\rho g_{lk}-\rho\partial_\rho g_{ij}\right)\Big|_{\rho=1/\epsilon},
\end{equation}
in which $L$ is the upper bound of the integral over $\tau$ and we have set $\rho=1/\epsilon$. After the subtraction of the counterterm 
\begin{equation}
    T^{{\rm ct}}_{ij}=-\frac{iL^2}{ 16\pi G_N}\; \frac{(d-1)g_{ij}}{\epsilon}+ \cdots,
\end{equation}
one then obtain the stress tensor 
\begin{equation}
    \langle T_{ij} \rangle=\frac{iL^2}{16\pi G_N}(g_{(2)ij}-g_{(0)ij}{\rm Tr}g_{(2)})\label{stress1}
\end{equation}
by taking the limit $\epsilon\rightarrow 0$. Moreover, with the help of the relation

\begin{equation}
    g_{(2)ij}=\frac{1}{2}(Rg_{(0)ij}+t_{ij}),\qquad {\rm Tr}\;t=-R,
\end{equation}
where $R=R\;[g_{(0)}]$ and $t_{ij}$ is a conserved symmetic tensor $\nabla^i t_{ij}=0$, we have

\begin{equation}
    \langle T_{ij} \rangle=\frac{iL^2}{32\pi G_N}\; t_{ij}.
\end{equation}
Therefore, after taking the trace, the Weyl anomaly is then deduced to be
\begin{equation}
    \langle T^i_i\rangle =-\frac{c}{24\pi}R,
\end{equation}
in which $c$ is the central charge on the celestial sphere
\begin{equation}
    c=\frac{i3L^2}{4G_N}.
\end{equation}
One can see the central charge will approach $i\infty$ as argued in \cite{Cheung:2016iub,Pasterski:2022lsl,Ogawa:2022fhy} if one treats $L$ as the scale of energy. 

The infinite behaviour of the central charge can be easily understood given the detail of the flat/CFT dictionary, which has been extensively studied in the case for scalar fields \cite{Hao:2023wln}. The onshell scalar field is studied by the separation of variables which splits the $\tau$ direction and other coordinates on the AdS hypersurfaces and it turns out that one could decompose a single scalar field into infinite number of modes labelled by the complex number $k$ making the scale dimension of the dual operator live on the principal series. Here we have the metric $G_{\mu\nu}$ while it is hard to apply the variable separation method to split the nonlinear Einstein equation into the $\tau$ dependent part and the other part which describes the equation of motion on the AdS hyperboloid therefore decompose the metric into the form of 
\begin{equation}
    G_{\mu\nu}(\tau,\rho,x)\longrightarrow G_{\mu\nu}(\rho,x;k)
\end{equation}
labelled by the parameter $k$. But here it is still reasonable to assume that the bulk metric is dual to infinite number of operators on the boundary described by stress-tensor modes denoted as $T_{ij}(x;k)$ and the energy-stress tensor calculated here are in fact the summation of all these modes. Each mode will contribute to the central charge in a finite way while the total effect will become infinite after summing over all the modes labelled by $k$ treated as the frequency space dual to the $\tau$ direction.
\subsection{Soft Sector}

Now, based on our study of hard sector, we move on to the study of soft sector. In order to obtain the next leading order correction $S^{1}_{\rm gr,ren}$, one needs to consider the higher order terms in $1/\tau$ of the onshell action \eqref{onshell}. 
 \begin{eqnarray} 
    && S^{0}_{\rm gr,reg}+S^{1}_{\rm gr,reg}=\frac{-1}{16\pi G_N}\int^{L}_0 d\tau \tau \int_{S^2}d^2x \sqrt{\hat{G}}\frac{4\rho}{\sqrt{{\hat{G}}^{\rho\rho}}}\big( (d+\rho g^{ij}\partial_\rho g_{ij})\\ &&\qquad\qquad\qquad\qquad\qquad\qquad\qquad\qquad-\frac{1}{\tau}(2\nabla^iA_i-g^{ij}\rho\partial_\rho\sigma_{ij} +\sigma^{ij}\rho\partial_\rho g_{ij})\big),\nonumber
 \end{eqnarray}
therefore regulated action at the first order now becomes
\begin{eqnarray}
    &&S_{{\rm gr, reg}}^{1}=\frac{L}{8\pi G_N}\int_{S^2}d^2z\; \sqrt{g_{(0)}} \; \rho\;\Big( 2\nabla^iA_i-\rho g^{ij}\partial_\rho \sigma_{ij}+\sigma^{ij}\rho\partial_\rho g_{ij}\\&&\qquad\qquad\qquad\qquad\qquad\qquad\qquad\qquad\qquad -(2m+2dM)(d+\rho g^{ij}\partial_\rho g_{ij})\Big)\Big|_{\rho=1/\epsilon} \nonumber
\end{eqnarray}
where the integral over $\tau$ has already been performed and the contribution in the second line comes from the determinant $\sqrt{\hat{G}}$ and the norm vector factor $\sqrt{G_{\rho\rho}}$. Together with the expression of the extrinsic curvature
\begin{equation}
    K_{ij}=-\rho(\rho\partial_\rho g_{ij}+g_{ij})+\frac{\rho}{\tau}\Big(\nabla_iA_j+\nabla_jA_j-\rho\partial_\rho \sigma_{ij}-\sigma_{ij}+2m(\rho\partial_\rho g_{ij}+g_{ij})\Big),
\end{equation}
one can obtain the soft stress tensor $T_{ij}^{1,{\rm reg}}$
\begin{eqnarray}
    &&T^{1,{\rm reg}}_{ij}=\frac{iL}{8\pi G_N}\rho\Big((\nabla_iA_j+\nabla_jA_i-\rho\partial_\rho\sigma_{ij}-\sigma_{ij})+(\rho g^{lk}\partial_\rho g_{lk}+d)\sigma_{ij}\\&& -(2\nabla_iA^i-\rho g^{lk}\partial_\rho \sigma_{lk}+\rho \sigma^{lk}\partial_\rho g_{lk})g_{ij}+2m(\rho\partial_\rho g_{ij}+g_{ij})-2m(\rho g^{lk}\partial_\rho g_{lk}+d)g_{ij}\Big)\nonumber
\end{eqnarray}
by checking the first order of $K_{ij}-\gamma_{ij}K$ then performing the integral over $\tau$ like we have done for the hard sector. From the above expression one can see the stress tensor will go to infinity at large $\rho$. One of the divergent term comes from $\nabla_i A_j$ while the other term comes from the first order metric on the sphere $\sigma_{ij}$. However, taking the constraint $A^{(0)}_i=A^{(2)}_i=0$ and the counterterm 
\begin{equation}
    S_{{\rm gr},{\rm ct}}=-\frac{d-1}{8\pi G_N}\int_0^Ld\tau \int_{S^2} d^2x \;\tau\sqrt{\hat{G}}\; \big(4\rho^2 \hat{G}_{\rho\rho}\big)^{-\frac{1}{2}}
\end{equation}
into consideration, we have
\begin{equation}
    T_{ij}^{\rm ct}=\frac{iL^2}{16\pi G_N}(1-d)\rho\big(g_{ij}+\frac{2}{L}(\sigma_{ij}-2mg_{ij})\big)+\cdots
\end{equation}
therefore the corresponding finite renormalised stress tensor at first order becomes
\begin{eqnarray}
   &&\langle T_{ij}^{1}\rangle=- \frac{iL}{8\pi G_N}\Big(({\rm Tr}\;\sigma^{(2)}-\sigma^{(0)}_{kl}\;g_{(2)}^{kl})\;g^{(0)}_{ij}+{\rm Tr}g^{(2)}\sigma_{ij}^{(0)}\nonumber\\ &&\qquad\qquad\qquad\qquad\qquad\qquad-\sigma^{(2)}_{ij}+2m^{(0)}(g^{(2)}_{ij}-g^{(0)}_{ij}{\rm Tr} g^{(2)})\Big).
\end{eqnarray}
\subsection{Ward Identities}
Given the flat/CFT dictionary and the specific expression of the energy-stress tensor, now we turn to study Ward identities concerning $\langle T_{ij}\rangle$ with the help of constraints on the gravity metric studied before. For Weyl anomaly, we will perform the calculation from the boundary point of view which means that the indices now are raised and lowered by the metric $\bar{G}_{ij}$. After taking the trace, for the soft stress tensor, we have\footnote{Here $\langle T^i_i \rangle^1$ represents the first order of $\bar{G}^{ij}\langle T_{ij}\rangle$, i.e $\langle T^i_i \rangle^1=\langle T_{ij}^1\rangle g_{(2)}^{ij}-\langle T^0_{ij}\rangle\sigma_{(0)}^{ij}$.}
 \begin{eqnarray}     
   && \langle T_i^i\rangle^{1}=-\frac{iL}{16\pi G_N}\Big(2(d-1)\big({\rm Tr} \sigma^{(2)}-\sigma^{(0)}_{ij}g_{(2)}^{ij}-2m^{(0)}{\rm Tr}g^{(2)}\big)\nonumber\\ &&\qquad\qquad\qquad\qquad\qquad+{\rm Tr}g^{(2)}{\rm Tr}\sigma^{(0)}-\sigma^{(0)}_{ij}g_{(2)}^{ij}\Big),
\end{eqnarray}
which is equivalent to
\begin{equation}
    \langle T^i_i \rangle^1=\frac{iL}{\pi G_N}m^{(2)}
\end{equation}
after using the relation \eqref{rel0}, \eqref{rel2} and setting $d=2$. As we have studied, $m^{(2)}$ could be an arbitrary scalar function therefore it will contribute to the  anomaly at the subleading order in an arbitrary way. At the same time, one should note that the form of anomaly are highly constrained in two dimensional conformal field theory \cite{Deser:1993yx,Bonora:1985cq}, the anomaly should be proportional to the Euler density, which means one should consider a special class of asymptotically flat spacetime in order to make the boundary field theory conformal. Here, we choose to consider a set of solutions of $m^{(2)}$ written as
\begin{equation}
    {\rm Tr}\sigma^{(2)}=\alpha M {\rm Tr} g^{(2)} \qquad dm^{(2)}=(1-\frac{\alpha}{4})M{\rm Tr}g^{(2)} \label{gauge}
\end{equation}
satisfying the constraint \eqref{rel0} and \eqref{rel1} for a real parameter $\alpha$. Such choice could be treated as the fix of gauge for soft sector like the gauge of asymptotic AdS hyperboloids are fixed in terms of Fefferman Graham coordinates where $G_{\rho\rho}^0=1/4\rho^2$. Therefore, we will have
\begin{equation}
    \langle T^i_i \rangle^{1}=-\frac{iL}{16\pi G_N}(2\alpha-8){\rm Tr}g^{(2)}m^{(0)}.\label{c1}
\end{equation}
In such case, we can treat the contribution from $m^{(2)}$ as part of the central charge at subleading order. To determine the central charge at the order of $1/L$, we can use the relation
    \begin{equation}
    \langle T^i_i\rangle =-\frac{c}{24\pi}R\;[\bar{G}]\label{c2}.
\end{equation}
where $R[\bar{G}]=R[g_{0}]+\frac{1}{L}(g^{ij}_{(0)}\delta R_{ij}-\sigma^{ij}_{(0)}R_{ij}[g_{(0)}])$. By checking the formula \eqref{c1} and \eqref{c2} specifically, we have  
\begin{equation}
     c=\frac{i3L^2}{4G_N}+\frac{iML3(\alpha-2)}{2G_N}
\end{equation}
from which we can see the central charge will have first order correction that depends on the geometry of spacetime characterised by the parameter $M$ and $\alpha$ while we leave higher order correction for further investigation.

Before going to study the conservation laws of energy-stress tensor, we first recall some lessons learnt from the AdS/CFT correspondence. For the asymptotically AdS case, the spacetime behaves like a box and no particle could finally reach the infinity while this fits the calculation that the dual CFT energy-stress tensor is conserved. Following our definition of asymptotic flatness, the hard sector is built up by the AdS hyperboloid therefore the dual energy-stress tensor on the celestial sphere is expected to be conserved at leading order, written as
\begin{equation}
    \bar{\nabla}^j\langle T_{ij}\rangle^{0}=0,
\end{equation}
which can be deduced using \eqref{rel2} and \eqref{stress1}. Comparing with the AdS spacetime, one of the main feature for asymptotically flat spacetime is that there could be gravitational radiation at the boundary and thus the system is not strictly closed i.e, energy could be carried in or away by particles passing through the null boundary. Now, for the stress tensor at subleading order, we have
\begin{eqnarray}
    &&\bar{\nabla}^j\langle T_{ij}\rangle^{1}=-\frac{iL}{16\pi G_N}\big(2\nabla_i({\rm Tr}\sigma^{(2)}-\sigma^{(0)}_{kl}g^{kl}_{(2)})-2\nabla^j\sigma^{(2)}_{ji}\nonumber \\ &&\qquad\qquad\qquad\qquad\qquad\qquad+\sigma_{(0)}^{mk}\nabla _m g^{(2)}_{ki}+\nabla^j{\rm Tr}g_{(2)}\sigma^{(0)}_{ij}\big)
\end{eqnarray}
where we have used the the relation \eqref{relation2} and the covariant derivative $\bar{\nabla}_i$ is with respect to the background metric $\bar{G}_{ij}$ of the boundary CFT. Moreover, taking the constraint \eqref{constraint4} into consideration,  we have
\begin{eqnarray}
    \bar{\nabla}^j\langle T_{ij} \rangle=\frac{iL}{16\pi G_N}(8\nabla_i m^{(2)}+4M\nabla_i{\rm Tr}g^{(2)}-\sigma^{mk}_{(0)}\nabla_m g^{(2)}_{ki}),
\end{eqnarray}
from which we can see that the stress tensor is not conserved at the subleading order due to the existence of soft modes therefore we can interpret such soft modes as the radiation modes which generate the flow of energy through the null boundary.
\section{Conclusion}
In this paper, we have developed the holographic renormalisation procedure for the gravitational action on the asymptotically flat background then obtain the flat/CFT dictionary between the $d+2$ dimensional theory in the bulk and the $d$ dimensional CFT on the celestial sphere. Based on the construction of flat/CFT dictionary, we then obtain a precise map between the asymptotic bulk data and the conformal energy stress tensor. By considering the conformal anomalies, we deduce the value of the central charge up to the first subleading order, which comes from the soft sector of the energy stress tensor. It turns out that the central charge takes complex value approaching infinity. Such behaviour has already been argued for long while here, given the flat/CFT duality relation, we may ask if the complex central charge implies that the boundary CFT is not unitary and we will leave this to further investigation. Moreover, we can see the value of central charge is expressed as power expansion according to the energy cut off $L$ and the physical implication of such expansion is also not clear from either bulk or boundary point of view. 

The introduction of the cuff off $L$ can be treated as the extra input when performing the renormalisation for the gravitational action. For scalar case \cite{Hao:2023wln}, we also have the integral over $\tau$ from zero to infinity while there is no need to introduce the cut off to make the action finite. That is because the integral over $\tau$ together with the weight function and $\tau$ modes will produce orthogonal relations telling us the coupling between different modes labelled by $k$. We have seen such orthogonal relation by solving Klein-Gordon equation explicitly while here the treatment for Einstein equation will be much harder as we have briefly discussed in section \ref{Dictionary}. In this paper, we have not studied each graviton mode in a microscopic way while we choose to study those infinite number of modes macroscopically by introducing the cut off $L$.

As we have seen, for the flat/CFT duality, most of the nontrivial results come from the contribution of the soft sector. It will lead to a non-conserved stress-energy tensor from boundary point of view. Such stress tensor makes the behaviour of boundary CFT more complicated while it enables us to investigate the gravitational bulk radiation from boundary point of view. Therefore the interpretation of non-conserved part of the energy-stress tensor is more like the introduction of heat bath or matter fields studied in the AdS/CFT correspondence.  

The definition of asymptotic flatness is clarified in the whole paper as \eqref{asmetric} while one may ask if we could consider the asymptotically flat spacetime in a more general sense. It is interesting to explore how the renormalisation works if Fefferman Graham gauge is broken. For example, one can consider the case that $G_{\rho\rho}^0$ takes arbitrary form or $G_{\rho i}^0 \neq 0$. For the spacetime in \eqref{asmetric}, the choice of spatial radius $\rho$ on the AdS hyperboloid as the IR regulator is straight forward since it will not break the asymptotic symmetry while the development of holographic renormalisation will become more complicated if one wants to deal with more general metric. 

For the soft sector, we also meet the similar problem like the gauge fixing of the AdS hyperboloid and this comes from the freedom of the choice of ${\rm Tr}\sigma^{(2)}$ or equivalently $m^{(2)}$. As we have seen, the trace part ${\rm Tr}\sigma^{(2)}$ tends to contribute to the subleading part of the anomalies of the stress tensor $\langle T^i_i \rangle$ in a arbitrary way while the form of Weyl anomaly is highly constrained from the CFT point of view. After the holographic renormalisation, to make the field theory coming from the bulk gravitational theory conformal, we have to further fix the gauge of ${\rm Tr}\sigma^{(2)}$ as shown in \eqref{gauge}. Here we have the freedom to do so but this leads to the problem that if all the asymptotically flat gravitational theory is dual to the CFT on the boundary. In fact, the definition of asymptotically flat spacetime is a vague concept in gravity as we have discussed in the introduction. In addition to the Ricci flat condition $R_{\mu\nu}=0$, one should also make the spacetime approach to flat at infinity so that recover enough flat space results and properties. On the other hand, the CFT is well studied thus such mismatching makes the  construction of flat/CFT duality challenging.

At the end, we illustrate some connections to the celestial holography. The duality between the bulk metric and the boundary stress-energy tensor has also been studied in the work \cite{Kapec:2016jld,Cheung:2016iub} in order to manifest the duality between bulk scattering soft theorems and boundary Ward identities. Here we derived the expression for the boundary energy-stress tensor with the guidance of the flat/CFT dictionary \eqref{dictionary} while this enables us to determine both of the hard and soft sector of the energy-stress tensor in terms of expansions from bulk metric in a local way therefore the boundary Ward identities will naturally correspond to the Einstein equation in the bulk. It is interesting to quantize those gravitational modes and find the relation between these two kinds of bulk gravity and boundary energy-stress tensor duality. We will leave this to further investigation.
\acknowledgments
I would like to thank my father Qinghe Hao and mother Xiulan Xu for providing funding for the tuition and accommodation fees when studying PhD at the University of Southampton (Fourth year tuition fee is covered by university funding for teaching). I also wish to thank Enrico Parisini for various insightful discussions on the structure of asymptotcially flat spacetime. I am grateful to  Kostas Skenderis and Marika Taylor for valuable comments on the manuscript. 
\appendix
\renewcommand{\thesubsection}{\Alph{subsection}}
\setcounter{section}{0}
\section{Asymptotic symmetries}
Given the Killing vector $\xi$, we can then further write down the variation of the metric as
\begin{equation}
    \mathcal{L}_{\xi}G_{\mu\nu}=\xi^\sigma \partial_\sigma G_{\mu\nu}+G_{\mu\sigma}\;\partial_\nu\xi^\sigma+G_{\nu\sigma}\;\partial_\mu\xi^\sigma,
\end{equation}
in which the expression is true for generic metric. In this article we will work in the Fefferman Graham gauge which means we have $G_{\tau\tau}=-1$ and $G_{\tau a}=0$. In such gauge, for $\tau\tau$ component, we have
\begin{equation}
    \mathcal{L}_{\xi}G_{\tau\tau}=\xi^\sigma\partial_\sigma G_{\tau\tau}+2G_{\tau\sigma}\;\partial_\tau\xi^\sigma=-2\partial_\tau\xi^\tau
\end{equation}
while for the $\tau\rho$ and $\tau z$ component we have
\begin{eqnarray}
\mathcal{L}_{\xi}G_{\tau\rho}&=&G_{\tau\sigma}\;\partial_\rho\xi^\sigma+G_{\rho\sigma}\;\partial_\tau\xi^\sigma \\
    &=& -\partial_\rho \xi^\tau +G_{\rho\rho}\partial_\tau \xi^\rho+ G_{\rho z}\partial_\tau \xi^z+G_{\rho \bar{z}}\partial_\tau \xi ^{\bar{z}}
\end{eqnarray}
and
\begin{eqnarray}
    \mathcal{L}_{\xi}G_{\tau z}&=&G_{\tau\sigma}\;\partial_z\xi^\sigma+G_{z \sigma}\;\partial_\tau\xi^\sigma \\
    &=& -\partial_z\xi^\tau+G_{\rho z}\partial_\tau \xi^\rho+G_{zz}\partial_\tau \xi^z+G_{z\bar{z}}\partial_\tau \xi^{\bar{z}}.
\end{eqnarray}
For the spatial $\rho z \bar{z}$ part, we have
\begin{equation}
    \mathcal{L}_\xi G_{z\bar{z}}=\xi^\sigma\partial_\sigma G_{z\bar{z}}+G_{z\sigma}\partial_{\bar{z}}\xi^\sigma+G_{\bar{z}\sigma}\partial_z\xi^\sigma
\end{equation}
\begin{equation}
    \mathcal{L}_\xi G_{zz}=\xi^\sigma\partial_\sigma G_{zz}+2G_{z\sigma}\partial_z \xi^\sigma
\end{equation}
\begin{equation}
    \mathcal{L}_\xi G_{\rho\rho}=\xi ^\sigma\partial_\sigma G_{\rho\rho}+2G_{\rho\sigma}\partial_\rho \xi^\sigma
\end{equation}
\begin{equation}
    \mathcal{L}_\xi G_{\rho z}=\xi^\sigma \partial_\sigma G_{\rho z}+G_{\rho\sigma}\partial_z\xi ^\sigma +G_{z\sigma}\partial_\rho \xi^\sigma.
\end{equation}
\section{Fefferman and Graham Coordinates} \label{ads}

In terms of Fefferman and Graham coordinates $(r,x_i)$.  The metric for asymptotic AdS spacetime takes the form
\begin{equation}
    \hat{G}_{rr}=\frac{1}{r^2}, \qquad \hat{G}_{ij}=\frac{g_{ij}}{r^2},\qquad \hat{G}_{r i}=0,
\end{equation}
in which the asymptotic behaviour is described by the function $g_{ij}(r,x_i)$. In such coordinates, the connection is then given by
\begin{equation}
    \Gamma_{rr}^r=-\frac{1}{r},\qquad \Gamma^r_{r i}=\Gamma^i_{rr}=0.
\end{equation}
For the other components that involve $r$  component, in terms of the function $g_{ij}$, the connections can be written as
\begin{equation}
    \Gamma^r_{ij}=-\frac{r^2}{2}\;\partial_r \hat{G}_{ij}=\frac{g_{ij}}{r}-\frac{1}{2}\partial_rg_{ij}
\end{equation}
and
\begin{equation}
    \Gamma^i_{r j}=\frac{1}{2}\hat{G}^{ik}\partial_r \hat{G}_{kj}=-\frac{\delta^i_j}{r}+\frac{1}{2}g^{ik}\partial_r g_{kj}
\end{equation}
For the connections that do not involve the $r$ component, they are determined by the function $g_{ij}$ and one can treat them as the connection of $g_{ij}$, i.e, $\Gamma^i_{jk}=\hat{\Gamma}^i_{jk}[g]$. 

Given the connections, we can use them to calculate the Ricci tensor following the definition \eqref{ricci}, the $rr$ component is given by
\begin{eqnarray}
     R_{rr}^{d+1}[\hat{G}]&=&\partial_r\Gamma^k_{r k}+\Gamma^l_{r k}\Gamma^k_{r l}-\Gamma^r_{rr}\Gamma^k_{kr}\nonumber\\
     &=&\frac{1}{2}\;g^{ij}\;\partial^2_r \;g_{ij}-\frac{1}{2r}g^{ij}\partial_r g_{ij}-\frac{1}{4}g^{ij}\; g^{lm}\;\partial_r g_{il} \;\partial_r g_{jm}+\frac{d}{r^2}
\end{eqnarray}
and the $ir$  components is determined to be
  \begin{eqnarray}
      R_{ir}^{d+1}[\hat{G}]&=&\partial_i\Gamma_{r k}^k-\partial_k\Gamma^k_{r i}+\Gamma^k_{r l}\Gamma^l_{ik}-\Gamma^m_{r i}\Gamma^k_{km} \\
      &=& \frac{1}{2}\partial_i(\;g^{lm}\;\partial_r g_{lm})-\frac{1}{2}\partial_k(\;g^{kl}\;\partial_r g_{il})+\frac{1}{2}g^{km}\;\Gamma^l_{ik}\;\partial_r g_{lm}-\frac{1}{2}g^{mk}\;\Gamma^l_{lm}\;\partial_r g_{ik}.\nonumber
  \end{eqnarray}
Moreover, in terms of covariant derivative $\nabla_i$ with respective to the metric $g_{ij}$,
$R_{ir}$ can be simplified to 
\begin{equation}
    R_{ir}^{d+1}[\hat{G}]=\frac{1}{2}\nabla_i(g^{lm}\;\partial_r g_{lm})-\frac{1}{2}\nabla^j \partial_r g_{ji}.
\end{equation}
The $ij$ component is given by
\begin{eqnarray}
&&R_{ij}^{d+1}[\hat{G}]=R_{ij}[g]  -\partial_r\Gamma^r_{ij}  +\Gamma^k_{ir}\Gamma^r_{jk}+\Gamma^r_{il}\Gamma^l_{jr}-\Gamma^r_{ij}\Gamma^l_{lr}-\Gamma^r_{ij}\Gamma^r_{rr}\\&&\qquad\quad
= R_{ij}[g]+\frac{1}{2}\partial^2_rg_{ij}+\frac{d}{r^2}g_{ij}+\frac{1-d}{2r}\partial_r g_{ij}-\frac{1}{2}g^{km}\partial_rg_{ki}\partial_rg_{mj}\nonumber\\&&\qquad\qquad+\frac{1}{4}\partial_rg_{ij}g^{lm}\partial_rg_{lm}-\frac{1}{2r}g_{ij}g^{lm}\partial_rg_{lm},\nonumber 
\end{eqnarray}
in which the induced Ricci tensor of $g_{ij}$ is denoted as $R_{ij}[g]$.

\section{Equation of Motion} \label{eom}
To calculate the Ricci tensor, we will use the convention
\begin{equation}
    R_{\mu\nu\rho}^{\;\;\;\;\;\;\sigma}=\partial_\mu\Gamma_{\nu\rho}^\sigma- \partial_\nu\Gamma_{\mu\rho}^\sigma+\Gamma_{\mu\lambda}^\sigma\Gamma^\lambda_{\nu\rho}-\Gamma_{\nu\lambda}^\sigma\Gamma^\lambda_{\mu\rho}
    \end{equation}
so that the tensor is given by
 \begin{equation}
        R_{\mu\nu}= R_{\mu\rho\nu}^{\;\;\;\;\;\;\rho}=\partial_\mu\Gamma_{\nu\rho}^\rho- \partial_\rho\Gamma_{\mu\nu}^\rho+\Gamma_{\mu\lambda}^\rho\Gamma^\lambda_{\nu\rho}-\Gamma_{\rho\lambda}^\rho\Gamma^\lambda_{\mu\nu}.\label{ricci}
    \end{equation}
For $d$ dimensional spacetime, in terms of the Milne coordinates, the Einstein equation at linear level could be written as
\begin{equation}
    R_{\tau\tau}[G]=\frac{4\rho^2}{\tau}\partial_\tau h_{\rho\rho}+2\rho^2\partial^2_\tau h_{\rho\rho}+\frac{1}{2}g^{ij}\partial^2_\tau h_{ij}+\frac{1}{\tau}g^{ij}\partial_\tau h_{ij}
\end{equation}
\begin{eqnarray}
    &&R_{\tau i}[G]=2\rho^2\nabla_i\partial_\tau h_{\rho\rho}+\frac{1}{2\rho}\nabla_i(\;g^{jm}\partial_\tau h_{jm})-\frac{1}{2\rho}\partial_\tau\nabla^kh_{ki}-2\rho^2\partial_\rho\partial_\tau h_{i\rho}\nonumber\\&&\qquad\qquad-4\rho\partial_\tau h_{i\rho}-2\rho^2\partial_\tau h_{\rho i}\Gamma^a_{a\rho}
\end{eqnarray}
in which we have
\begin{equation}
   \Gamma^a_{a\rho}= \frac{d-2}{2}\frac{1}{\rho}+\frac{1}{2}g^{lm}\partial_\rho g_{lm}
\end{equation}
and $R_{\tau \rho}$ is deduced to be
\begin{eqnarray}
  &&  R_{\tau \rho}[G]=-\frac{1}{2\rho^2}g^{ij}\partial_\tau h_{ij}+\frac{1}{2\rho}\partial_\rho(g^{ij}\partial_\tau h_{ij})-2\rho\partial_\tau h_{\rho\rho}-\frac{1}{2\rho}\partial_\tau\nabla^i h_{i\rho}\nonumber\\&& \qquad\qquad\qquad-2\rho^2\partial_\tau h_{\rho\rho}\Gamma^a_{a\rho}+\frac{1}{2\rho}g^{ij}\partial_\tau h_{jk}\Gamma^k_{i\rho}.
\end{eqnarray}
For other components, with the help of the expansion of the Ricci curvature, 
\begin{equation}
    R_{ab}^{d+1}[\hat{G}^{(0)}+h]=R_{ab}^{d+1}[\hat{G}^{(0)}]+\frac{1}{2}(\hat{\nabla}^2h_{ab}+\hat{\nabla}_a\hat{\nabla}_b h-\hat{\nabla}_c\hat{\nabla}_ah^c_b-\hat{\nabla}_c\hat{\nabla}_bh^c_a)+\mathcal{O}(h^2)
\end{equation}
where $\hat{\nabla}_a$ is the covariant derivative with respect to the metric $\hat{G}^{(0)}_{ab}$. Then one can obtain
\begin{eqnarray}
    &&R_{\rho \rho}[G]=-dh_{\rho\rho}-\frac{d+2}{2}\tau\partial_\tau h_{\rho\rho} -\frac{1}{2}\tau^2 \partial^2_\tau h_{\rho\rho}-\frac{\tau}{8\rho^3}g^{ij}\partial_\tau h_{ij}+(4\rho^2\Gamma^i_{\rho j}\Gamma^j_{i\rho}-\frac{g^{ij}}{\rho^2}\Gamma^\rho_{ij}\nonumber\\&&\qquad\qquad+\frac{g^{ij}}{\rho}\Gamma^k_{i\rho}\Gamma^\rho_{jk}-4(d+\rho g^{ij}\partial_\rho g_{ij}))h_{\rho\rho}-(2\rho(d+\rho g^{ij}\partial_\rho g_{ij})+\frac{g^{ij}}{2\rho}\Gamma^\rho_{ij})\partial_\rho h_{\rho\rho}\nonumber\\&&\qquad\qquad+\frac{1}{2\rho}\nabla^i\nabla_ih_{\rho\rho} -\frac{1}{\rho}\nabla_i\partial_\rho h_\rho^i-\frac{2}{\rho}\Gamma^k_{i\rho}\nabla^ih_{k\rho}-\frac{1}{\rho^2}\nabla_ih^i_{\rho}\nonumber
    +\frac{1}{2\rho}\partial^2_\rho h^i_i -\frac{1}{2\rho^2}\partial_\rho h^i_i\nonumber\\&&\qquad\qquad+\frac{1}{\rho}\Gamma^i_{\rho j}\partial_\rho h^j_i+\frac{1}{2\rho^3}h^i_i+\frac{1}{\rho}g^{ij}\Gamma^k_{i\rho}\Gamma^l_{j\rho}h_{kl}-\frac{1}{\rho^2}\Gamma^i_{\rho j}h^j_i-\frac{1}{\rho}\Gamma^i_{\rho j}\Gamma^k_{\rho i}h^j_k
\end{eqnarray}
where we have $h^i_j=g^{ik}h_{kj}$ and $h^i_\rho=g^{ik}h_{k\rho}$. The connections used here are given by  $ \Gamma^{\rho}_{\rho\rho}=-\frac{1}{\rho}$, $\Gamma^i_{j\rho}=\frac{\delta^i_j}{2\rho}+\frac{1}{2}g^{ik}\partial_\rho g_{kj}$ and 
    $\Gamma^\rho_{ij}=-2\rho^2 g_{ij}-2\rho^3 \partial_\rho g_{ij}$.
 For the components of $R_{i\rho}$, $R_{ij}$ we have
\begin{eqnarray}
    &&R_{i\rho}[G]=-\frac{d+1}{2}\tau\partial_\tau h_{i\rho}-\frac{1}{2}\tau^2\partial^2_\tau h_{i\rho}-dh_{i\rho}-2\rho\nabla_i h_{\rho\rho}-2\rho^2\nabla_k(\Gamma^k_{i\rho} h_{\rho\rho})-\frac{1}{2\rho}\nabla^m(\Gamma^\rho_{mi}h_{\rho\rho})\nonumber
    \\&&\qquad\qquad-2\rho^2\Gamma^a_{a\rho}\nabla_i h_{\rho\rho}+\frac{1}{2\rho}(\nabla^m\nabla_mh_{i\rho}-\nabla_k\nabla_ih^k_\rho)+2\rho^2\partial_\rho(\Gamma^j_{\rho i}h_{\rho j})-\frac{1}{2\rho}\partial_\rho(\Gamma^\rho_{ij}h^j_{\rho})\nonumber\\&& \qquad \qquad+\frac{3}{2}\Gamma^\rho_{mi}\Gamma^k_{n\rho}g^{mn}h_{k\rho}+\frac{g^{mn}}{2\rho}\Gamma^\rho_{mn}\Gamma^j_{\rho i}h_{j\rho} +\frac{1}{2\rho}g^{mn}\Gamma^k_{m\rho}\Gamma^\rho_{nk}h_{i\rho}+\frac{1}{2\rho}\Gamma^j_{k\rho}\Gamma^{\rho}_{ij}h^k_\rho\nonumber\\&& \qquad\qquad-\frac{1}{2\rho^2}g^{mn}\Gamma^\rho_{mn}h_{i\rho}+2\rho^2\Gamma^j_{k\rho}\Gamma^k_{j\rho}h_{\rho i}
    +\Gamma^a_{a\rho}(-2\rho h_{\rho i}+2\rho^2\Gamma^j_{\rho i}h_{\rho j}+2\rho^2\Gamma^k_{i\rho}h_{\rho k}-\frac{1}{2\rho}\Gamma^\rho_{ij}h^j_\rho)\nonumber\\&&\qquad\qquad +8\rho\Gamma^j_{i\rho}h_{\rho j}-2h_{i\rho}+\frac{1}{\rho}\Gamma^\rho_{ij}\partial_\rho h^j_\rho -2\rho\partial_\rho h_{i\rho} -2\rho^2\Gamma^a_{a\rho}\partial_\rho h_{\rho i}
       +\frac{1}{\rho}\nabla_k(\Gamma^l_{i\rho}h^k_l)\nonumber\\&&
       \qquad\qquad-\frac{1}{2\rho}\Gamma^k_{m\rho}\nabla^m h_{ik}+\frac{1}{2\rho}\Gamma^k_{j\rho}\nabla_i h^j_k-\frac{1}{2\rho}\nabla^j(\Gamma^k_{j\rho}h_{ik}) -\frac{1}{2\rho}\nabla_j(\Gamma^j_{\rho k} h^k_i)+\frac{1}{2\rho}\Gamma^j_{k\rho}\nabla_jh^k_i\nonumber \\&&\qquad\qquad+\frac{1}{2\rho^2}\nabla_kh^k_i-\frac{1}{2\rho^2}\nabla_i h-\frac{1}{2\rho}\Gamma^k_{i\rho}\nabla_k h+\frac{1}{2\rho}\nabla_i\partial_\rho h -\frac{1}{2\rho}\nabla_j\partial_\rho h^j_i
\end{eqnarray}
and 
\begin{eqnarray}
    &&R_{ij}[G]=-dh_{ij}-\frac{d+1}{2}\tau\partial_\tau h_{ij}-\frac{1}{2}\tau^2\partial_{\tau}^2h_{ij}-\frac{\tau}{2}g_{ij}(g^{mn}\partial_\tau h_{mn}+4\rho^3\partial_\tau h_{\rho\rho})
    \nonumber\\&& \qquad\qquad+\partial_\rho(4\rho^2\Gamma^\rho_{ij}h_{\rho\rho}) -2\rho^2(\Gamma^k_{\rho i}\Gamma^\rho_{k j}h_{\rho\rho}+\Gamma^k_{\rho j}\Gamma^\rho_{ki}h_{\rho\rho})+\frac{1}{\rho}\Gamma^\rho_{ni}\Gamma^\rho_{nj}g^{mn}h_{\rho\rho} +2\rho^2\nabla_i\nabla_jh_{\rho\rho}\nonumber\\
   && \qquad\qquad+4\rho^2\Gamma^a_{a\rho}\Gamma^{\rho}_{ij}h_{\rho\rho}-\Gamma^\rho_{ij}\partial_\rho(2\rho^2h_{\rho\rho})-2\partial_\rho(\rho^2\nabla_ih_{\rho j})-2\partial_\rho(\rho^2\nabla_jh_{\rho i}) \nonumber\\
   && \qquad\qquad -2\rho^2\nabla_k(\Gamma^k_{i\rho} h_{j\rho})-2\rho^2\nabla_k(\Gamma^k_{j\rho}h_{i\rho})+\frac{1}{2\rho}\Gamma^\rho_{kj}\nabla_i h^k_\rho+\frac{1}{2\rho}\Gamma^\rho_{ki}\nabla_j h^k_{\rho}+\frac{1}{\rho}\nabla_k(\Gamma^\rho_{ij}h^k_\rho)\nonumber\\&& \qquad\qquad-2\rho^2\Gamma^a_{a\rho}(\nabla_ih_{j\rho}+\nabla_jh_{i\rho})-2\rho^2\Gamma^k_{\rho i}h_{\rho j}-2\rho^2\Gamma^k_{\rho j}h_{\rho i}-2\rho^2\Gamma^k_{\rho j}\nabla_i h_{k\rho}-2\rho^2\Gamma^k_{\rho i}\nabla_j h_{k\rho}\nonumber \\&& \qquad\qquad-\frac{1}{2\rho}\nabla^n(\Gamma^\rho_{ni}h_{\rho j})-\frac{1}{2\rho}\nabla^n(\Gamma^\rho_{nj}h_{i\rho})-\frac{1}{2\rho}\Gamma^\rho_{mi}\nabla^m h_{\rho j}-\frac{1}{2\rho}\Gamma^\rho_{mj}\nabla^m h_{\rho i}\nonumber
   \\&& \qquad\qquad -\frac{1}{2\rho}(\nabla_k\nabla_i  h^k_j+\nabla_k\nabla_j h^k_i)+\frac{1}{2\rho}\nabla_i\nabla_j h+\frac{1}{2\rho}\nabla_m\nabla^m h_{ij}+2\rho^2\partial^2_\rho h_{ij}+2\rho\partial_\rho h_{ij}\nonumber
    \\&& \qquad\qquad-2\rho^2\big(\partial_\rho(\Gamma^k_{\rho i}h_{kj})+\partial_\rho(\Gamma^k_{\rho j}h_{ki})+\Gamma^k_{\rho i}\partial_\rho h_{kj}+\Gamma^k_{\rho j}\partial_\rho h_{ki}\big)+\frac{1}{2\rho^2}\Gamma^\rho_{ij}h-\frac{1}{2\rho}\Gamma^\rho_{ij}\partial_\rho h\nonumber\\&&\qquad\qquad +\frac{1}{2}\Gamma^\rho_{ki}\partial_\rho(\frac{1}{\rho}h^k_j)+\frac{1}{2}\Gamma^\rho_{kj}\partial_\rho(\frac{1}{\rho}h^k_i)-\frac{1}{2}\partial_\rho(\frac{1}{\rho}\Gamma^\rho_{ki}h^k_j)-\frac{1}{2}\partial_\rho(\frac{1}{\rho}\Gamma^\rho_{kj}h^k_i)\nonumber
   \\&&\qquad\qquad +2\rho^2(\Gamma^k_{\rho i}\Gamma^m_{\rho k}h_{mj}+\Gamma^k_{\rho j}\Gamma^m_{\rho k}h_{mi}+2\Gamma^k_{\rho i}\Gamma^m_{\rho j}h_{mk})-\frac{1}{2\rho}(\Gamma^\rho_{ki}\Gamma^m_{\rho j}+\Gamma^\rho_{kj}\Gamma^m_{\rho i})h^k_m\nonumber\\ &&\qquad\qquad+\frac{1}{2\rho}(\Gamma^\rho_{ki}\Gamma^k_{\rho m}h^m_j+\Gamma^\rho_{kj}\Gamma^k_{\rho m}h^m_i+\Gamma^k_{\rho i}\Gamma^\rho_{km}h^m_j+\Gamma^k_{\rho j}\Gamma^\rho_{km}h^m_i)-2\rho\Gamma^k_{\rho i}h_{k j}-2\rho\Gamma^k_{\rho j}h_{k i}\nonumber\\ && \qquad\qquad-\frac{1}{2\rho}\Gamma^a_{a\rho}(\Gamma^\rho_{im}h^m_j+\Gamma^\rho_{jm}h^m_i) 
   +\frac{g^{mn}}{2\rho}\Big(\Gamma^\rho_{mn}(\Gamma^k_{i\rho}h_{kj}+\Gamma^k_{\rho j}h_{ki})+\Gamma^k_{n\rho}(\Gamma^\rho_{mj}h_{ki}\nonumber\\ &&\qquad\qquad+\Gamma^\rho_{mi}h_{kj})-\Gamma^\rho_{mn}\partial_\rho h_{ij} \Big).
\end{eqnarray}

Given the Einstein tensor, then one should be able to deduce $h_{ab}$ exactly by solving the Einstein equation $R_{\mu\nu}=0$. For simplicity, we consider the equation at the leading order which means for $h_{ab}$ we have
\begin{equation}
    h_{\rho\rho}=\frac{1}{\rho^2\tau}m(\rho,z,\bar{z}),\qquad h_{ij}=\frac{\rho}{\tau}\sigma_{ij}(\rho,z,\bar{z}),\qquad h_{\rho i}=\frac{1}{\tau}A_i(\rho,z,\bar{z})
\end{equation}
where higher order terms of $1/\tau$ are omitted.




 \bibliographystyle{JHEP}
\bibliography{ref.bib}

\end{document}